# Causal Analysis of Shapley Values: Conditional vs. Marginal

Ilya Rozenfeld

Capital One Financial

9/5/2024

## Abstract

Shapley values, a game theoretic concept, has been one of the most popular tools for explaining Machine Learning (ML) models in recent years. Unfortunately, the two most common approaches, conditional and marginal, to calculating Shapley values can lead to different results along with some undesirable side effects when features are correlated. This in turn has led to the situation in the literature where contradictory recommendations regarding choice of an approach are provided by different authors.

In this paper we aim to resolve this controversy through the use of causal arguments. We show that the differences arise from the implicit assumptions that are made within each method to deal with missing causal information. We also demonstrate that the conditional approach is fundamentally unsound from a causal perspective. This, together with previous work in [1], leads to the conclusion that the marginal approach should be preferred over the conditional one.

## 1. Introduction

With the increase of popularity of Machine Learning (ML) models, the explainability of their prediction became an important and active area of research. Starting with a seminal paper of Lundberg and Lee [2] the **SH**apley **A**dditive ex**P**lanation (SHAP) has become one of the most popular methods for local explanations[1].

SHAP's rapid increase in popularity was mainly due to three reasons. First, the method relies on the well-established concepts from game theory [3]. Second, unlike earlier work on Shapley values [4, 5], [2] proposed a fast model-agnostic computational method; Kernel SHAP, which made computation of Shapley values for big data ML models feasible[2]. Lastly, the method was implemented in open-source code[3], which continues to be actively maintained with many features and great visualization capabilities.

As a part of Shapley values calculations, the model with some of the features removed needs to be evaluated [6]. The two most popular approaches are marginal and conditional which are also referred to as interventional and observational, respectively. They produce the same results when features are uncorrelated but differ for models with correlated features while exhibiting some anomalous behaviors. This has led to contradictory recommendations in the literature [1, 7-11].

---

[1] Determining features' contributions to model's prediction for a single observation.
[2] Faster methods have been developed since then such as TreeShap [18].
[3] https://github.com/shap/shap

In this paper we seek to resolve the controversy by extending causal analysis in [1] to a conditional approach. We demonstrate that the differences between the two approaches are due to how they implicitly account for the missing causal information. We show that a conditional approach assumes causation based on correlation, violating the most basic principles of statistics, and therefore, should be avoided.

## 2. Background

### 2.1. Shapley Values

SHAP method is based on theory of cooperative games [3], an exploration of how group of players work together to achieve a payout. More precisely, given a set of players $\mathcal{M} = \{1, \ldots, M\}$ and a measure of payoff that can be achieved by cooperating players in coalition $S$ as value function $v(S)$ defined for all possible $S \subseteq \mathcal{M}$, the goal is to determine $j^{th}$ player contribution $\phi_j(v)$ to the total payoff $v(\mathcal{M})$.

To make the attributions fair, four axioms are usually imposed:

1. **Efficiency**: Total payout is fully distributed across players
$$v(\mathcal{M}) = \sum_{j \in \mathcal{M}} \phi_j(v)$$
2. **Dummy:** A player that doesn't contribute to any coalition receives zero attribution
$$v(S \cup \{j\}) = v(S), \forall S \implies \phi_j(v) = 0$$
3. **Symmetry**: Two players that contribute equally to every coalition get the same attributions
$$v(S \cup \{j\}) = v(S \cup \{i\}), \forall S \implies \phi_j(v) = \phi_i(v)$$
4. **Additivity**: Sum of the player's attributions from two different games is equal to attribution from combined game
$$\phi_i(v_1 + v_2) = \phi_i(v_1) + \phi_i(v_2)$$

It turns out that with these axioms the attribution problem has unique solution

$$\phi_j(v) = \sum_{S \subseteq \mathcal{M} \setminus \{j\}} \frac{|S|! \, (M - |S| - 1)!}{M!} \left( v(S \cup \{j\}) - v(S) \right) \qquad (3.1)$$

where $|S|$ denotes the number of players and $\mathcal{M} \setminus \{j\}$ a set of all players excluding $j^{th}$ one. The formula for $\phi_j(v)$ can be viewed as the weighted sum of the marginal contributions $v(S \cup \{j\}) - v(S)$ of $j^{th}$ player. The quantities $\phi_j(v)$ are referred to as Shapley values.

### 2.2. Shapley Values for ML models

Suppose we have an ML model $f(X)$ with a set of features $X = \{X_1, \ldots, X_M\}$. Additionally, suppose we want to explain the model's prediction at some sample $x^*$. This sample is sometimes referred to as explicand. To apply cooperative game theory to model explanation, three components need to be identified: total payout, players, and value function.

Since we are interested in the local explanations, prediction $f(x^*)$ is defined as the total payout. Features $X$ are players with complimentary subsets $X_S$ and $X_{\bar{S}}$ being as in-coalition and out-of-coalition, respectively. Then from Efficiency axiom the prediction $f(x^*)$ can be decomposed into Shapley values as

$$f(x^*) = \phi_0 + \sum_{j=1}^{M} \phi_j(x^*) \tag{3.2}$$

where $\phi_0$ is the reference value commonly defined as $E[f(X)]$

Defining the value function $v(S)$ is not as straightforward. The reason is that the model needs to be evaluated on the in-coalition subset of features $x_S^*$ while it requires the full set $x^*$. Among different approaches [6], the two by far most popular are the conditional and marginal averages over out-of-coalition features. The value function with conditional average is defined as

$$v(S) = E[f(X)|X_S = x_S^*] \tag{3.3}$$

and with marginal as

$$v(S) = E[f(\{x_S^*, X_{\bar{S}}\})] \tag{3.4}$$

The two approaches are illustrated in Figure 1 on the dataset with three features with the first one in coalition and the last two out of coalition. The explicand is the first observation. The conditional approach restricts the averaging to the samples where $X_S = 1$, the value of $X_S$ in the explicand (red bracket). In the marginal approach, on the other hand, the averaging is over all samples (green bracket) but with all the values of in-coalition feature set to 1 (green box).

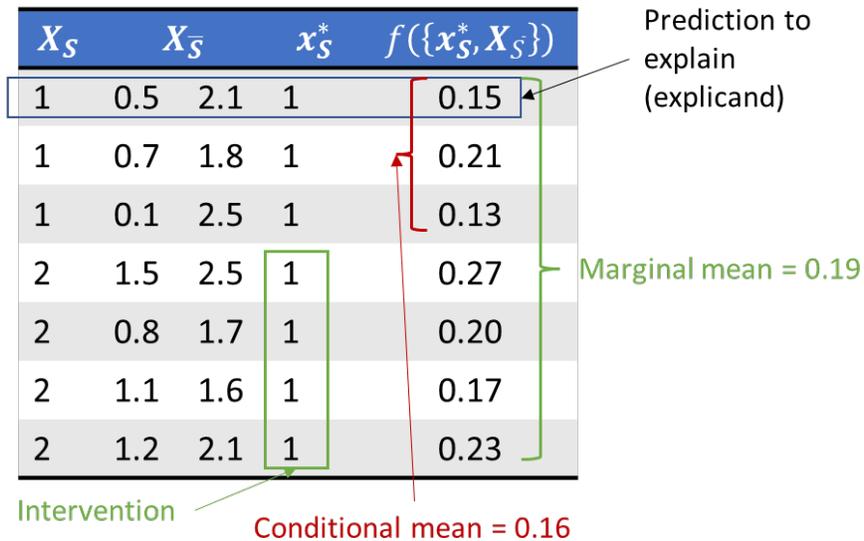

Figure 1: Illustration of conditional and marginal approaches to the out-of-sample features removal.

## 2.3. Issues with correlated features

When features are uncorrelated both approaches produce the same results. However, for correlated features not only would the two approaches arrive at different results but would also lead to some undesirable effects. These issues are well known and were discussed at great length throughout the literature (see [12], for example). We describe them briefly here for completeness.

First, we illustrate the issue with the conditional approach using a simple linear model with two correlated features

$$f(X) = \beta_0 + \beta_1 X_1 + \beta_2 X_2 \qquad (3.5)$$

with $E[X_1] = E[X_2] = 0$, $E[X_1^2] = E[X_2^2] = 1$ and $E[X_1 X_2] = \rho$. In this case, it is fairly simple to estimate Shapley values as

$$\phi_1 = \beta_1 x_1^* + \frac{\rho}{2}(\beta_2 x_1^* - \beta_1 x_2^*)$$
$$\phi_2 = \beta_2 x_2^* + \frac{\rho}{2}(\beta_1 x_2^* - \beta_2 x_1^*) \qquad (3.6)$$

While the first terms in each equation of Eqs. (3.6) are intuitive, the second ones are difficult to interpret. However, the main problem with Eqs. (3.6) appears when it is assumed that feature $X_1$ has no impact on the model i.e. $\beta_1 = 0$. Then $\phi_1 = \rho/2 \, \beta_2 x_1^*$ contrary to the expected $\phi_1 = 0$[4].

Next, we consider marginal averaging. The scatter plot of training data with correlated features $X_1$ and $X_2$ is shown in Figure 2 (blue circles) along with explicand (red dot). The samples on which the model would need to be evaluated for marginal averaging are shown by green dots. A portion of evaluation samples lies well outside the sample on which the model would've been trained forcing it to extrapolate into regions with little or no training data[5]. In fact, some feature combinations may not be physically possible. For example, if two features $X_1$ and $X_2$ were age and binary variable denoting possession of driver's license (1 – Yes, 0 – No), respectively, the combinations like $X_1 < 16$ and $X_2 = 1$ would be impossible.

---

[4] While this seems to violate the Dummy axiom, it is important to keep in mind that the axioms are defined relative to the value function $v(S)$ and not the model $f(X)$; therefore, there is no contradiction.
[5] Per [19] this issue is relevant to any metric that utilizes permutations, not just Shapley Value.

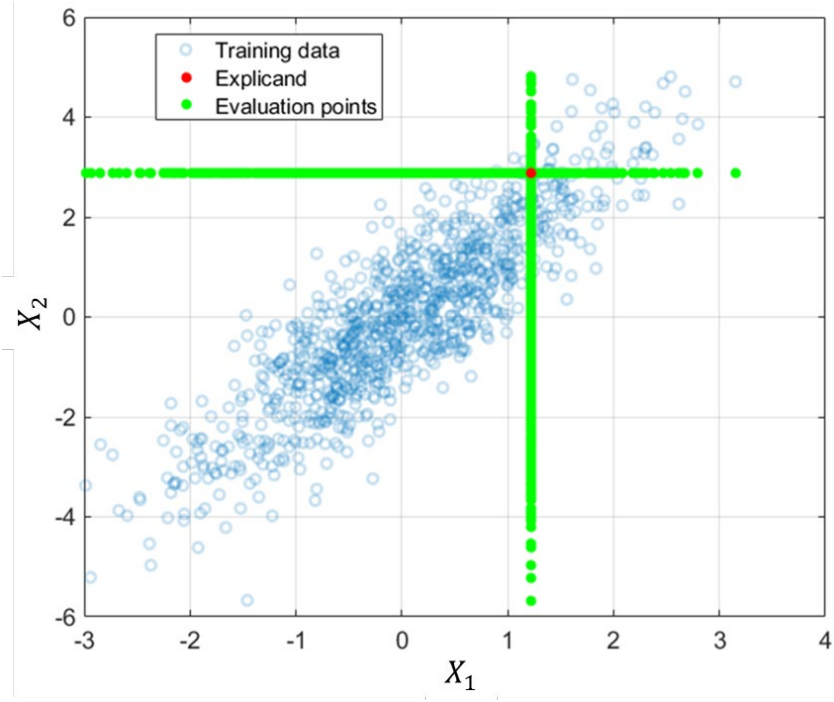

Figure 2: Scatter plot of correlated features $X_1$ and $X_2$ (blue circles) with evaluation samples (green dots) used in marginal averaging for a given explicand (red dot).

This has led to disagreements among the researchers as to what method should be used. Some authors argue in favor of the marginal approach [1, 7, 8], while the others argue for the conditional one [9, 10]. In [11] the authors propose that both approaches could be used with the choice based on context. The marginal approach should be used when the researcher's goal is to explain the model ("true to the model"), and the conditional approach should be used when the goal is to explore relationships within the data ("true to the data").

An interesting alternative is to utilize causality in calculating Shapley values. [13] and [14] propose to adjust weights of marginal contributions based on at least partially specified causal structure. Wang et. al [15] introduce an approach called Shapley Flow where attributions are assigned to graph's edges instead of nodes. These approaches can capture indirect impacts in addition to direct ones at the expense of violating the Symmetry axiom.

## 3. Marginal vs. Conditional from Causal Perspective

Janzing, et al. [1] propose a causal argument to justify the use of marginal averaging. They draw a distinction between Directed Acyclical Graphs (DAG).[6] for generating process leading to outcome $Y$ from causes $\widetilde{X}$ (Figure 3, panel (a)) and DAG for generating model prediction $\hat{Y}$ from features $X$ that are inputs into model $f$ (Figure 3, panel (b)) which themselves stem from causes $\widetilde{X}$. Then they establish that marginal averaging is equivalent to do-operator from Pearl's do-calculus [16],

---

[6] For background on causal inference and DAGs we refer reader to [16].

$$E[f(\{x_S^*, X_{\bar{S}}\})] = E[f(X)|do(X_S = x_S^*)]. \qquad (4.1)$$

The do-operator in Eq. (4.1) denotes intervening on features $X_S$ by setting them to $x_S^*$. In DAG this operation removes all arrows coming into corresponding nodes as illustrated in panel (c) of Figure 3 where the intervention on features $X_1$ and $X_2$ is depicted. This removes backdoor paths, and with them correlations, between intervened features to all other ones. Note that the dependencies between non-intervened features remain intact. The equivalence of marginal averaging and do-operator provides causal justification for the former.

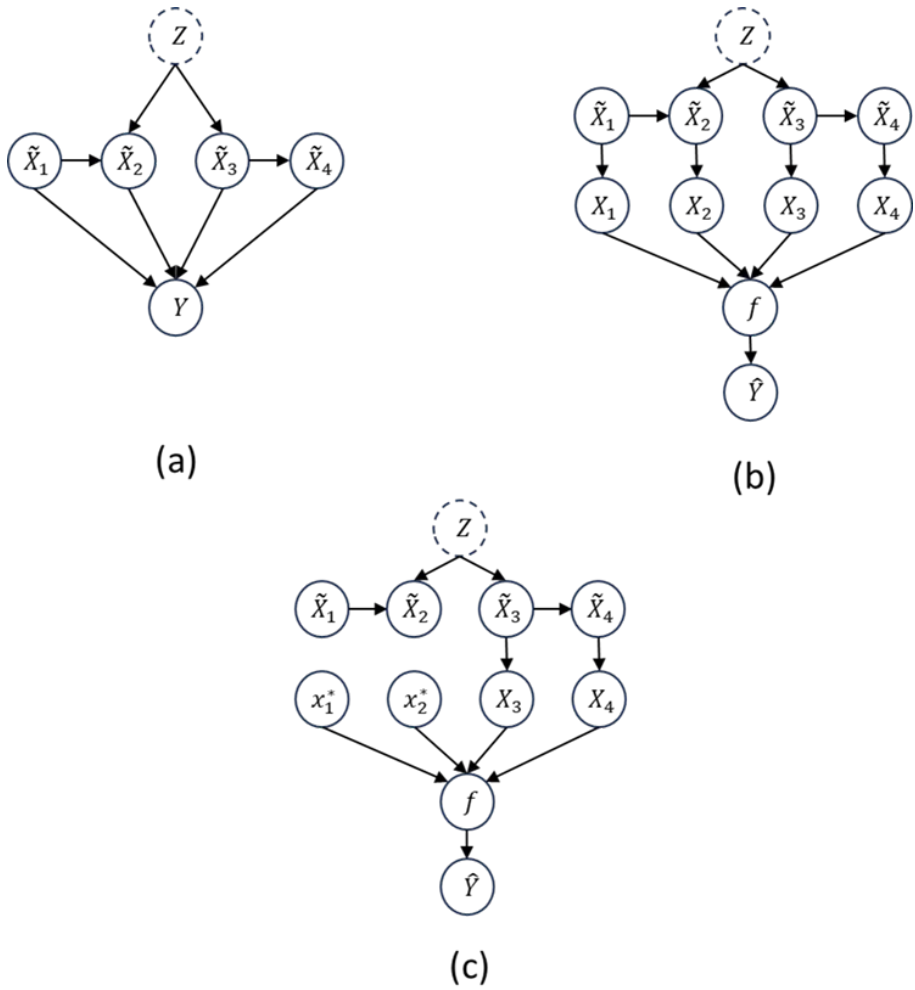

Figure 3: (a) DAG for generation of outcome $Y$ from causes $\tilde{X}$; (b) DAG for generating model prediction $\hat{Y}$ from features $X$; (c) DAG in panel (b) after intervening on features $X_1$ and $X_2$. Nodes with solid and dashed boundaries are observable and latent (i.e. unobservable) causes, respectively.

Next, we demonstrate the relationship between conditional and causal Shapley values [13, 14]. For this we utilize some of the results from [13] and [14] where causal Shapley values were derived for some basic DAGs. In particular, the causal Shapley values for right-directed chain and fork DAG (panel (a) in Figure 4) are

$$\phi_1^r = v(\{X_1\}) - v(\emptyset)$$
$$\phi_2^r = v(\{X_1, X_2\}) - v(\{X_1\})$$
(4.2)

where $v(S) = E[f(X)|X_S = x_S^*]$. Similarly, the causal Shapley values for the right-directed chain and fork DAG (panel (b) in Figure 4) are

$$\phi_1^l = v(\{X_2, X_1\}) - v(\{X_2\})$$
$$\phi_2^l = v(\{X_2\}) - v(\emptyset).$$
(4.3)

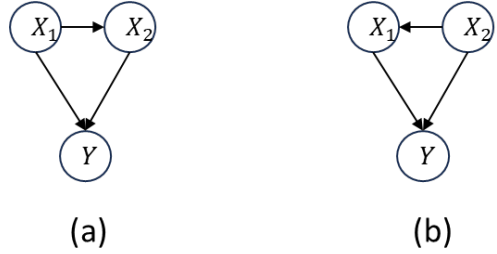

Figure 4: (a) Right-directed chain and fork; (b) left directed chain and fork.

Averaging Eqs. (4.2) and (4.3) we arrive at

$$\frac{1}{2}(\phi_1^r + \phi_1^l) = \frac{1}{2}(v(\{X_1\}) - v(\emptyset)) + \frac{1}{2}(v(\{X_2, X_1\}) - v(\{X_2\}))$$
$$\frac{1}{2}(\phi_2^r + \phi_2^l) = \frac{1}{2}(v(\{X_1, X_2\}) - v(\{X_1\})) + \frac{1}{2}(v(\{X_2\}) - v(\emptyset))$$
(4.4)

These expressions are the same as those for conditional Shapley values since conditional averaging was used in calculating causal Shapley values. To understand this surprising result, we apply the framework from [1] and consider what happens when we intervene with conditional distribution as shown in Figure 5 for the example of confounder DAG at the top. As in Figure 3 the distinction is made between causes $\widetilde{X}$ and features $X$. When intervening on feature $X_1$ (bottom left DAG in Figure 5), the backdoor path is eliminated. However, conditioning on $X_1$ impacts distribution of $X_2$ creating direct causal path from $X_1$ to $X_2$ (orange arrow) and right-directed chain and fork DAG. Similarly, when intervening on $X_2$ (bottom right DAG in Figure 5), a direct causal path from $X_2$ to $X_1$ (orange arrow) is created leading to left-directed chain and fork DAG.

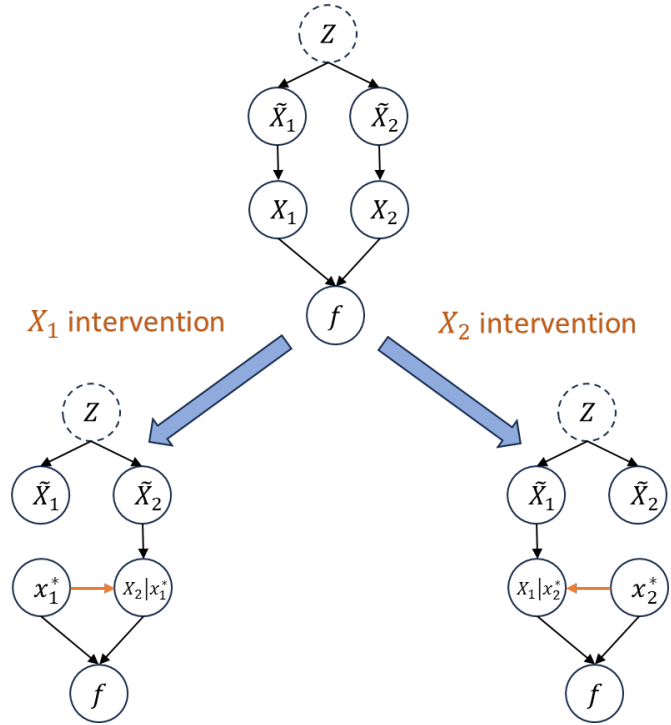

Figure 5: Intervening with conditional averaging.

It follows then that conditional averaging introduces non-existent causal effects based on the presence of correlation in the data, equating correlation with causation in violation of the fundamental principle of statistics. Moreover, the direction of these effects changes depending on the feature being intervened. While we illustrated these effects on the example of the model with two features, the conclusions apply to the models with any number of features. In this case the bi-directional causal connections will be created between each pair of correlated features and then averaged together. This implies that conditional approach is fundamentally flawed and shouldn't be used in the presence of correlated features.[7]

## 4. Conclusions

The issues encountered in both approaches to Shapley values are not accidental. Ultimately, explainability is about causality. After all, we are trying to determine how a certain prediction was produced by a model. Pearl in [17] calls data "profoundly dumb" because it contains information only about correlations and not causations. By extension, the model is not aware of the existence or directions of causal relationships between the features either. It is this lack of the causal information, and not anything specific to framework or calculation methods, that led to the issues with Shapley values discussed in this paper. It is likely that any explainability method relying solely on model and data will suffer from some kind of problems. They may manifest in different ways, but they will be present.

In [12], the use of causal information for Shapley values in [13] and [14] was seen as a limitation and a source of presumably unnecessary complexity. However, as was shown here when no explicit causality structure is provided, it will be replaced by implicit assumptions. We would argue that, on the contrary, the attempts to incorporate causal information, even if through assumptions, should be welcomed. In fact, we would consider the ability of the Shapley values framework to incorporate causal reasoning is one of its major strengths compared to other methods.

---

[7] This doesn't apply to conditional averaging in causal Shapley values calculations, like in [13] and [14], since non-real causal paths are handled through adjustment of weights of marginal contributions.

The difference between marginal and conditional approaches' results and their respective problems is due to essentially opposite ways the problem of missing causal information is handled. In the marginal approach all the unknown causal dependencies between the features are ignored and concentration is on only known causal information i.e. direct paths from features to model prediction. On the other hand, in the conditional approach, the Shapley value is a superposition of all possible pairwise causal connections, real or fantom, between the features.

While issues for both approaches stem from the same cause, missing causal information, their nature is very different. The conditional approach is fundamentally flawed violating one of the fundamental statistical principles. Conversely, the marginal approach, is sound from a theoretical perspective and has strong causal justification. The issues with the marginal approach are either due to sparsity of the data or the model's poor extrapolation properties. These are more of a practical nature and therefore can often, at least partially, be remedied, even for the cases with data with impossible features' values combinations. We plan to explore the impacts of extrapolation in marginal approach and ways to deal with them in our next paper.